\documentclass[aps,prl,floatfix,superscriptaddress,twocolumn]{revtex4}
\usepackage{graphics}
\begin{document}

\title{Ultrafast deterministic generation of entanglement 
in a time-dependent asymmetric two-qubit-cavity system} 

\author{Alexandra Olaya-Castro}\email{a.olaya@physics.ox.ac.uk}
\author{Neil F. Johnson}\email{n.johnson@physics.ox.ac.uk}
\affiliation{Centre for Quantum Computation and Department of Physics,
University of Oxford, Clarendon Laboratory, Parks Road, OX1 3PU, U.K}

\author{Luis Quiroga}\email{lquiroga@uniandes.edu.co}
\affiliation{Departamento de F\'{\i}sica, Universidad de Los Andes,
A.A.4976, Bogot\'a D.C., Colombia}

\date{\today}

\begin{abstract} 
We present an efficient scheme for the controlled generation
of pure two-qubit states possessing {\it any} desired degree of entanglement and
a {\em prescribed} symmetry in two cavity QED based systems,
namely, cold trapped ions and flying atoms. This is achieved via on-resonance
ion/atom-cavity couplings which are time-dependent and asymmetric, leading to a trapping
vacuum state condition which does not arise for identical couplings. A duality in the 
role of the coupling ratio yields states with a given
concurrence but opposing symmetries. The experimental feasibility of the proposed
scheme is also discussed.
\end{abstract}

\maketitle
Entanglement is one of the most subtle but striking quantum phenomena.
In the quantum information processing (QIP) field, it is considered 
as a physical resource which can be quantified
and manipulated. The power of this quantum resource has been demonstrated in
various applications such as state teleportation, quantum cryptography and
quantum dense coding \cite{rwnature02}. A crucial ingredient for any QIP protocol
is the {\em controlled} and {\em accurate} generation of entangled multi-qubit
states. A fast and deterministic generation process is highly desirable in
order to minimize decoherence effects. 

Strongly coupled ion/atom-photon systems are prime candidates for QIP \cite{rwnature02}. 
Many  schemes have been proposed for entangling atomic qubits using cavity
QED, with some even being implemented experimentally \cite{plenio99,
guan-guo00, osnaghi01}. Zheng $et$ $al.$ proposed entanglement via
a virtually excited cavity mode \cite{guan-guo00} which was later experimentally
realized \cite{osnaghi01}. Specifically, two atoms crossing a non-resonant
cavity are entangled by coherent energy exchange. 
However, {\em resonant} cavity QED is likely to offer faster
entanglement schemes. Plenio $et$ $al.$ considered singlet-state
generation through continuous observation of the cavity field in resonance with
two two-level atoms \cite{plenio99}.  The resulting entanglement
generation is, however, comparatively slow and probabilistic, requiring efficient active
measurements. Most theoretical studies on cavity-assisted
entanglement assume identical qubit-cavity couplings.
Nevertheless, it is known that spatial and/or temporal variations of the ion/atom-field coupling arise 
experimentally \cite{walther01,eschner01,duan03}. 
Zhou $et$ $al.$ recently considered entanglement within a
non-symmetric two-atom system coupled to a thermal cavity state 
\cite{q-guo03}, and found that the entanglement
is maximum for a particular static asymmetry in the coupling. However the full
consequences of having dynamical qubit-cavity couplings, are unknown. Indeed, most existing proposals are aimed at the
generation of maximally entangled two-qubit states without explicit
reference to the resulting state's symmetry. This motivates us to explore
systems in which the interplay between entanglement and state symmetry can be
studied directly, with a view to extracting these states for QIP {\em and}
to gain insight into physical mechanisms for entangling  large numbers of
quantum subsystems.

Here we propose a scheme to generate and control two-qubit
entangled states possessing a definite symmetry which can be chosen {\em a
priori}. Our results are obtained using a generalized Dicke model with
time-{\em dependent} qubit-cavity couplings, thereby mimicking actual
experimental conditions \cite{walther01,eschner01,duan03}. The resulting 
diversity of coupling profiles and timescales, offers a natural mechanism for
manipulating the collective qubit dynamics and hence controlling
the entanglement. We show that it is the geometric mean of the individual
couplings which dominates the entanglement's dynamical evolution. Exploiting a
particular trapping vacuum condition
\cite{weidinger99} in which the cavity-qubit state becomes separable but the
two-qubit state remains entangled, we show how full control 
of the quantum correlations can be
achieved. We then demonstrate the deterministic generation of a pure
two-qubit state with {\it any} degree of entanglement and a {\it prescribed}
symmetry. Our scheme has the following advantages:
(i) since it is resonance-based, the entanglement generation requires a
significatively shorter operation time than both the ion/atomic dipole decay
time and the radiation decay time; (ii) the resulting generation of pure states
with an arbitrary degree of entanglement, is deterministic and does not require
final state projection -- hence it is accurate; (iii) since the
system is non-symmetric with respect to qubit exchange, the
final-state symmetry can be controlled by choosing which qubit should 
initially be in its excited state. This final-state symmetry can therefore be
used to encode any quantum information present in the initial atomic
state.

We investigate the dynamical evolution of the concurrence
$C(\rho_{a,a})$ which measures the degree of two-qubit entanglement
\cite{wooters98}, and the two-qubit correlation function
$\langle \sigma_2^+ \sigma_1^-\rangle$ which measures the degree of
qubit exchange symmetry. For a singlet state
$\langle
\sigma_2^+ \sigma_1^-\rangle=-1/2$, while for a maximally-entangled triplet 
state $\langle \sigma_2^+ \sigma_1^-\rangle=1/2$. Hence negative values for
$\langle \sigma_2^+ \sigma_1^-\rangle$  are associated with an
anti-symmetric-like behavior while positive values are related to a
symmetric-like behavior.  Furthermore, we also investigate the qubit-field correlations
$\langle \sigma_i^+ a\rangle$ and find relations between these quantum
correlations and the qubit-field entanglement. 

We consider two possible scenarios: (i) cold trapped ions inside a high-finesse cavity 
where the Lamb-Dicke localization limit is assumed \cite{semiao02}
and (ii) flying two-level atoms crossing a cavity \cite{hennrich00}. 
In both cases, given that the atomic transitions are on resonance with the single-mode cavity
field, the Hamiltonian in the interaction picture and rotating-wave
approximation becomes ($\hbar=1$)
\begin{eqnarray} H_I=\sum_{i=1,2}f_i(t_i,t,\tau_i)\{a^\dag
\sigma_i^{-}+\sigma_i^{+}a\}
\label{Eq:ham}
\end{eqnarray} where $\sigma_i^{+}=|e_i\rangle \langle g_i|$,
$\sigma_i^{-}=|g_i\rangle \langle e_i|$ with $|e_i\rangle$ and $|g_i\rangle$
$(i=1,2)$ being the excited and ground states of the $i$'th qubit. Here $a^\dag$
and
$a$ are respectively the creation and annihilation operators for the cavity
photons. The time-dependent coupling of the cavity field with the $i$'th qubit,
which is injected at $t_i$ and interacts during a time $\tau_i$, is given by a
time-window function,
\begin{eqnarray} f_i(t_i,t,\tau_i)=[\Theta (t -\tau_i)-\Theta (t
-\tau_i-t_i)]\gamma_i(t)
\end{eqnarray} where $\gamma_i(t)$ is the time-dependent qubit-field
coupling strength. We focus here on the situation in which the qubits interact
simultaneously with the cavity mode such that $t_1=t_2=0$ and
$\tau_1=\tau_2=\tau$.
 
In the case of identical time-dependent couplings
$f_1(t_1,t,\tau_1)=f_2(t_2,t,\tau_2)=f(0,t,\tau)$, the Hamiltonian is symmetric
with respect to qubit exchange and hence commutes with the pseudo-spin operator
$J^2=J_x^2+J_y^2+J_z^2$ where $J_{x,y,z}$ are the usual angular momentum
operators. The dynamical evolution of the state does not mix different
$J-$sectors. $H_I$ commutes with itself at different times
as well as with the operator $V=\sum_{i=1,2}\{a^\dag
\sigma_i^{-}+\sigma_i^{+}a\}$. The combined qubit-field system follows a
unitary time evolution generated by the operator
$U=exp(-i\phi(t)V)$, where
$\phi(t)=\int_{0}^{t}\gamma(t')dt'$ represents the net effect of the
time-dependence in the qubit-field interaction.
For different but time-independent couplings, i.e. $\gamma_1\neq \gamma_2$, the
Hamiltonian has two important properties. First, it no longer commutes 
with $J^2$ and second, it is non-symmetric with respect to qubit-exchange. We
will show how these features can be exploited to control the total pseudo-spin
and hence the symmetry of the two-qubit state.  The time-evolution of the
system is more complex for the case of different but time-dependent couplings,
since generally the Hamiltonian does not commute with itself at different times. Hence a
history-dependent dynamics arises, governed by
$U=Texp(-i\int_{0}^{t}H_I(t')dt')$ where $T$ denotes the
time-ordering operator.
In all these cases, the number of excitations ${\cal N}=a^\dag a +
\sum_{i=1,2} \sigma^{+}_i \sigma^{-}_i$ is a conserved quantity. This implies
separable dynamics within subspaces having a prescribed eigenvalue
$N$ of ${\cal N}$. For $N=1$ a basis is given by
$\{|e_1,g_2,0\rangle,|g_1,e_2,0\rangle,|g_1,g_2,1\rangle\}$, while for
$N\ge 2$ a basis is given by $\{|e_1,e_2,N-2\rangle, |e_1,g_2,N-1\rangle,
|g_1,e_2,N-1\rangle, |g_1,g_2,N\rangle\}$. The third label denotes
the number of photons. For simplicity, we will restrict ourselves to consider
$|\Psi(0)\rangle=|e_1,g_2,0\rangle$ as the initial state so that the system's
dynamical evolution is confined to the single excitation subspace.
The whole system's quantum state at any time can be expressed as 
{\small \begin{eqnarray} |\Psi(t)\rangle=a_1(t)|e_1,g_2, 0\rangle +
a_2(t)|g_1,e_2,0\rangle +a_3(t)|g_1,g_2, 1\rangle
\end{eqnarray}}
Clearly the cavity mode acts as a third qubit, hence the single-qubit-field
entanglement can be quantified by the concurrence qubit$_{1,2}$-field as well. We
have found simple relations between these quantities:
$C(\rho_{a,a})(t)=2|\langle
\sigma_2^+(t) \sigma_1^-(t)\rangle|=2|a^*_1(t)a_2(t)|$,
$C(\rho_{a1,f})(t)=2|\langle \sigma_1^+(t) a(t)\rangle|=2|a^*_1(t)a_3(t)|$ and
$C(\rho_{a2,f})(t)=2|\langle \sigma_2^+(t) a(t)\rangle|=2|a^*_2(t)a_3(t)|$.
We consider the situation in which $\gamma_1(t)=r\gamma_2(t)$, where $r$ is a
constant. The unitary evolution is given by 
\begin{eqnarray} a_1(t)&=&1+ra_2(t)\nonumber \\ a_2(t)&=&-2\alpha
{\rm Sin}^2[\theta(t)/2]
\label{eq:coef}\\ a_3(t)&=&-i\sqrt{r \alpha} {\rm Sin}[\theta(t)]\nonumber
\end{eqnarray} where $\theta(t)=\int_0^t \omega(t')dt'$ is the effective vacuum
Rabi angle. The time-dependent frequency of the collective qubit mode coupled
to the cavity field is given by $\omega^2(t)=\gamma_1(t)^2 +\gamma_2(t)^2$, and
$\alpha=\gamma_1(t)\gamma_2(t)/\omega^2(t)=r/(1+r^2)$ denotes the relative
geometric mean of the couplings.

\begin{figure}
\resizebox{8cm}{!}{\includegraphics*{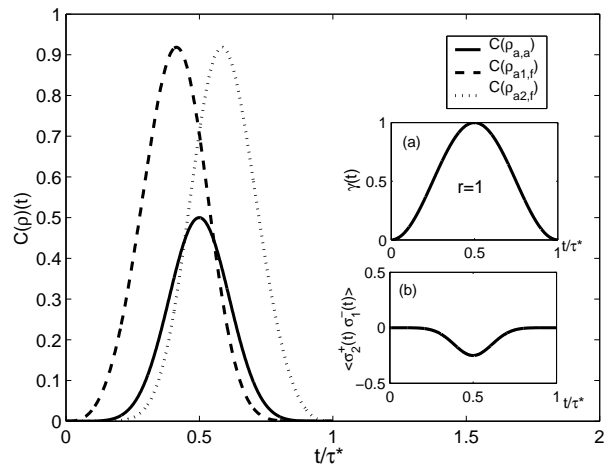}}
\caption{Two-qubit concurrence $C(\rho_{a,a})$, qubit$_1$-field concurrence
$C(\rho_{a1,f})$ and qubit$_2$-field concurrence $C(\rho_{a2,f})$ as a function
of time, for identical time-dependent couplings $r=1$. The initial state is
$|e_1,g_2,0\rangle$. Inset (a): Coupling strength as a function of time. Inset
(b): Time variation of the two-qubit correlation $\langle\sigma_2^+\sigma_1^-\rangle$.}
\label{fig:g1eqg2}
\end{figure}

To help understand the physics for non-symmetric
couplings, we express  $|\Psi(t)\rangle$ as 
\begin{eqnarray} |\Psi(t)\rangle=A(t)|\Phi^-,0\rangle + B(t)|\Phi^+, 0\rangle +
a_3(t)|g_1,g_2, 1\rangle
\label{eq:state}
\end{eqnarray} where $|\Phi^{\pm}\rangle=[|e_1,g_2\rangle \pm
|g_1,e_2\rangle]/\sqrt{2} $, $A(t)=[a_2(t)-a_1(t)]/\sqrt{2}$, and
$B(t)=[a_2(t)+a_1(t)]/\sqrt{2}$.  From Eqs. (\ref{eq:coef}) and (\ref{eq:state}) it
is clear that when $\theta(\tau^*)=\pi$ the qubit-field state is separable and
is given by $|\Psi(\tau^*)\rangle=[A(\tau^*)|\Phi^-\rangle +
B(\tau^*)|\Phi^+\rangle]\otimes|0\rangle$.  Therefore, at this special time a
trapping vacuum state condition holds, i.e. the cavity photon number is unchanged.
For identical couplings, i.e. $r=1$, the state $|\Phi^-,0\rangle$ is an
eigenstate of $H_I(t)$, thus the coefficient $A(t)$ remains constant in time,
i.e. $A(t)=A(0)=-1/\sqrt{2}$. Additionally for $t=\tau^*$,
$B(\tau^*)=1/\sqrt{2}$ such that the system's state becomes fully separable,
i.e. $|\Psi(\tau^*)\rangle=-|g_1,e_2,0\rangle$, hence there is no entanglement
in the qubit subsystem nor between the qubits and the field. 
Figure \ref{fig:g1eqg2} shows the dynamical evolution of the two-qubit
concurrence and the qubit-field concurrences, for the case of identical
time-dependent couplings. The initially
excited qubit becomes entangled with the cavity field faster than the initially
non-excited qubit.  Nevertheless, there is a symmetry in the dynamics of
individual qubit-field entanglements -- both qubits become identically
entangled with the field but at different times. At
$t=\tau^*/2$ when the individual qubit-field entanglements are identical, i.e. 
$C(\rho_{a1,f})=C(\rho_{a2,f})$, the two-qubit concurrence reaches its maximum
value $C(\rho_{a,a})(\tau^*/2)=1/2$ corresponding to the state
$|\Psi(\tau^*/2)\rangle=[-|\Phi^-,0\rangle -i|g_1,g_2, 1\rangle]/\sqrt{2}$.
Hence for $r=1$, two-qubit entanglement requires the existence of
qubit-field entanglement as well.

\begin{figure}
\resizebox{8cm}{!}{\includegraphics*{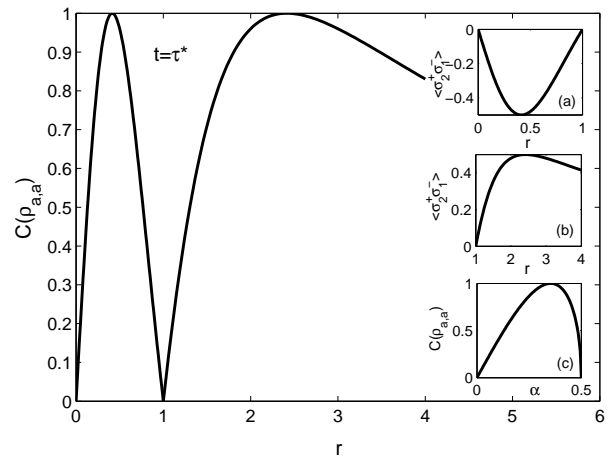}}
\caption{Two-qubit concurrence $C(\rho_{a,a})$ as a function of the coupling
ratio $r$, for $t=\tau^*$ and initial state $|e_1,g_2,0\rangle$. Inset (a):
Two-qubit correlation function $\langle\sigma_2^+(\tau^*)
\sigma_1^-(\tau^*)\rangle$  for $r<1$ and inset (b): $\langle\sigma_2^+(\tau^*)
\sigma_1^-(\tau^*)\rangle$  for $r>1$. Inset (c): $C(\rho_{a,a})$ as a function
of the relative geometric mean of the couplings $\alpha$.}
\label{fig:caavsr}
\end{figure}

With different couplings, the symmetry with
respect to qubit exchange is broken. Hence the coefficients $A(t)$ and
$B(t)$ can be fully controlled by choosing adequate values for the coupling
ratio $r$. In particular at $t=\tau^*$, any allowed superposition of
$|\Phi^-\rangle$ and
$|\Phi^+\rangle$ is obtainable in such a way that any prescribed two-qubit
entangled state can be achieved. Figure \ref{fig:caavsr} shows the two-qubit
concurrence as a function of $r$. Two important physical
consequences follow from Figure \ref{fig:caavsr}.  First, the symmetry 
of the two-qubit state is controlled by the parameter $r$. If the initially
excited qubit is interacting with the cavity through the weakest coupling ($r\le
1$), the two-qubit state is in the anti-symmetric sector
($\langle\sigma_2^+(\tau^*)
\sigma_1^-(\tau^*)\rangle < 0$ (Fig.\ref{fig:caavsr}(a)); otherwise it is in
the symmetric sector ($\langle\sigma_2^+(\tau^*) \sigma_1^-(\tau^*)\rangle \ge
0$ (Fig.\ref{fig:caavsr}(b)). Second, when the trapping vacuum condition is met
($t=\tau^*$), it is always possible to get two different matter states with
identical concurrences: one has a positive correlation ($r>1$) while the
other has a negative correlation ($1/r$). However, these two two-qubit states
have the same value of the relative geometric mean of the couplings
$\alpha=r/(1+r^2)$. Note that $\alpha\le1/2$, where the equality holds for
the identical coupling case. Therefore for any given value of $\alpha$, 
two-qubit states with identical concurrence but different symmetries can be found.
In particular, $r=\sqrt 2 +1$ and $r=\sqrt 2-1$ yield the same $\alpha=\frac
{1}{2\sqrt{2}}$ for which $C(\rho_{a,a})=1$. Hence two maximally entangled
states can be generated from a single excitation: the symmetric one for the
former value of $r$ and the fully anti-symmetric one (singlet state) for the
latter value of $r$.  Thus, $\alpha$ is the relevant
control parameter for generating a two-qubit state with any prescribed degree
of entanglement. In Fig.\ref{fig:caavsr}(c) the two-qubit
concurrence is shown as a function of $\alpha$. It is a
non-monotonic function of $\alpha$ as given by $C(\rho_{a,a})(\alpha)=4\alpha\sqrt{1-4\alpha^2}$.

\begin{figure}
\resizebox{8cm}{!}{\includegraphics*{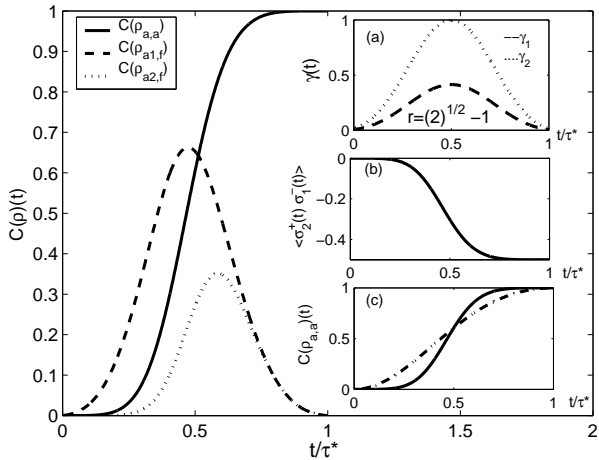}}
\caption{Dynamics of formation of the singlet state $|\Phi^-,0\rangle$. 
Two-qubit concurrence $C(\rho_{a,a})$, qubit$_1$-field concurrence $C(\rho_{a1,f})$ 
and qubit$_2$-field concurrence
$C(\rho_{a2,f})$ as a function of time, for  $r=\sqrt 2-1$ . The initial state
is $|e_1,g_2,0\rangle$. Inset (a): Coupling strengths as a function of time.
Inset (b): Time variation of the two-qubit correlation function $\langle\sigma_2^+ \sigma_1^-\rangle$. Inset (c): $C(\rho_{a,a})$ for different but
time-independent couplings (dashed-dotted line) and for the time-dependent
couplings as depicted in inset (a) (solid line). In both cases $r=\sqrt 2-1$ and the time $\tau^*$ is identical.}
\label{fig:singlet}
\end{figure}

\begin{figure}
\resizebox{8cm}{!}{\includegraphics*{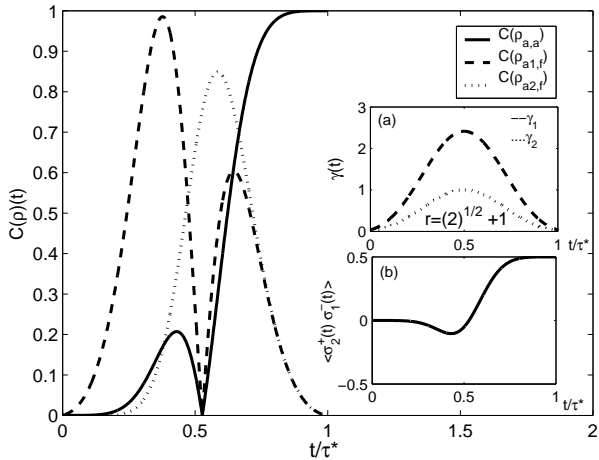}}
\caption{Dynamics of formation of a triplet maximally entangled state 
$|\Phi^+,0\rangle$. $C(\rho_{a,a})$, $C(\rho_{a1,f})$,
$C(\rho_{a2,f})$, and initial state as in Fig. 3, but now $r=\sqrt 2+1$. Inset (a): Coupling
strengths as a function of time. Inset (b): Time variation of the two-qubit correlation 
$\langle\sigma_2^+\sigma_1^-\rangle$.}
\label{fig:triplet}
\end{figure}

Figures \ref{fig:singlet} and \ref{fig:triplet} show the dynamics of formation
of maximally-entangled qubit states with different time-dependent couplings.
The two-qubit concurrence and qubit-field concurrences are shown as a
function of time for
$r=\sqrt 2 -1$ (Fig.\ref{fig:singlet}) and $r=\sqrt 2+1$
(Fig.\ref{fig:triplet}). In the formation of the singlet, the state remains
within a negative region of symmetry (Fig.\ref{fig:singlet}(b)), while for the
$r>1$ case the degree of symmetry evolves from negative to
positive (Fig.\ref{fig:triplet}(b)). In both cases, the maximal two-qubit
concurrence is reached at the moment when both individual qubit-field concurrences vanish.
It is worth noting that the time formation for the anti-symmetric state is
always larger than the time required for obtaining the symmetric one. For
example, for the time-dependent couplings depicted in
Fig.\ref{fig:singlet}(a) and given by
$\gamma_1(t)=\gamma_{max}sin^2(\pi t/\tau^*)$, the formation time
for the anti-symmetric state is
$\tau^*_{as}=(2\pi/\gamma_2^{max})\sqrt{\alpha/r}$ and the time required to
obtain the corresponding symmetric entangled state is $\tau^*_{s}=r\tau^*_{as}$
(N.B. $r<1$). Although a similar dynamics can also be observed with time-independent
couplings by keeping the same coupling ratio $r$ (trapping condition time 
$\tau^*=\pi/\omega$),
it is interesting to note that time-varying couplings yield to a
{\em switching} behavior of the two-qubit concurrence dynamics which could be important
for minimizing decoherence effects (Fig.\ref{fig:singlet}(c)).

In the experimental set-up described in Ref. \cite{walther01}, 
scanning of an optical field is performed by translating the
cavity assembly relative to the trapped ions' position.
Therefore, deterministic control of the spatial/temporal variation of the
ion-field coupling is feasible.  This experimental situation is 
likely  to generate couplings of order $\gamma_{max}=2\pi \times 3.9$MHz for calcium ions
(D$_{3/2}$ to P$_{1/2}$) with a dipole decay rate $\Gamma=2\pi \times 1.7$MHz 
and a cavity-photon decay rate $\kappa=2\pi \times 0.24$MHz \cite{wolfgang03}. 
The interaction time required to generate a symmetric maximally 
entangled state ($r=\sqrt{2}+1$) is $\tau_{s}^*=10^{-7}s$, which is 
one order of magnitude shorter than the photon decay time 
$T_r=1.3\times10^{-6}s$. We note that the required time to 
obtain an entangled state with $C(\rho_{a,a})=0.48$ ($\alpha=0.12$), 
in the symmetric region $(r=8)$, is even shorter: $\tau_s^*=3\times10^{-8}s$.  
Our proposal could also be realized with flying atoms sent simultaneously through a resonant cavity, 
in such a way that they follow different paths hence 
yielding different temporal coupling profiles.
Off-resonance coherent control of the collision of two Rydberg atoms has been experimentally 
implemented \cite{osnaghi01}. 
For these atomic Rydberg states, with principal 
quantum numbers $n=51$ $(|e\rangle)$ and $n=50$ $(|g\rangle)$, 
interacting with a cavity of quality factor $Q=7\times 10^7$ and 
corresponding photon lifetime $T_r=2 \times 10^{-4}s$, 
couplings of order $\gamma_{max}=2\pi \times 50$KHz were achieved. 
In this experimental scenario 
we estimate $\tau_{s}^*=10^{-5}s$ for $r=\sqrt{2}+1$ 
and $\tau_{s}^*=2\times 10^{-6}s$ for $r=8$. These times are 
obviously much shorter than the photon decay time. 
The flying-atoms scheme opens up the intriguing possibility of producing a stream of atomic pairs, where each one has a different entanglement and/or symmetry.
 
In summary, we have proposed a scheme to generate pure
two-qubit states with an arbitrarily prescribed degree of entanglement and
symmetry. Unlike previous cavity-based
schemes, the present one takes advantage of spatial and temporal variations
in the ion/atom-field coupling. It is achievable within very short operation
times since the qubits are on resonance with the cavity field. 
In particular, our scheme is realizable with 
present experimental methods and hence opens up the 
prospect of real-time engineering of multi-qubit entanglement in
asymmetric time-varying ion/atom-cavity systems.

We acknowledge funding from 
the Clarendon Fund and ORS (AOC), DTI-LINK (NFJ), and COLCIENCIAS project 1204-05-13614 (LQ).


\begin{thebibliography}{}
\bibitem{rwnature02} C.Monroe, Nature {\bf 416}, 238 (2002). 

\bibitem{plenio99} M.B. Plenio et al., Phys.Rev.A {\bf 59} 2468 (1999).

\bibitem{guan-guo00} S.-B. Zheng et al., Phys.Rev.Lett. {\bf 85}
2392 (2000).

\bibitem{osnaghi01} S. Osnaghi et al., Rev. Lett. {\bf 87}, 037902 (2001).

\bibitem{walther01} G.R. Guth\"ohrlein et al., Nature {\bf 414}, 49 (2001). 

\bibitem{eschner01} J. Eschner et al., Nature {\bf 413}, 495 (2001).

\bibitem{duan03} L.-M. Duan et al.,  Phys.Rev. A {\bf 67} 032305 (2003).

\bibitem{q-guo03} L.Zhou et al.,  quant-ph/0308086.

\bibitem{weidinger99} M.Weidinger et al., Phys.Rev.Lett {\bf 82}, 3795 (1999). 

\bibitem{wooters98} W.K. Wootters, Phys.Rev.Lett {\bf 80}, 2245 (1998).

\bibitem{semiao02} F.L.Semiao et al., Phys.Rev. A {\bf 66} 063403 (2002).

\bibitem{hennrich00} M.Hennrich et al., Phys.Rev.Lett {\bf 85}, 4872 (2000).

\bibitem{wolfgang03} M.Keller et al., J.Phys. B{\bf 36}, 613 (2003);
W.Lange (private communication).


\end{thebibliography}
\end{document}